\documentclass[useAMS,usenatbib,a4paper]{mn2e}

\usepackage{aas_macros}
\usepackage{graphicx}
\usepackage[usenames]{color}

\title{Discovery of a single faint AGN in a large sample of $z>5$ Lyman break galaxies}

\author[L. S.~Douglas et~al.]{Laura S.~Douglas$^{1}$, Malcolm N.~Bremer$^{1}$, Elizabeth R.~Stanway$^{1}$, Matthew D.~Lehnert$^{2}$\\
$^1$H H Wills Physics Laboratory, Tyndall Avenue, Bristol, BS8 1TL, UK\\
$^2$Laboratoire d'Etudes des Galaxies, Etoiles, Physique et Instrumentation GEPI, Observatoire de Paris, Meudon, France}

\begin{document}

\date{Accepted 2007 January 25. Received 2007 January 24; in original form 2006 December 18 }

\pagerange{\pageref{firstpage}--\pageref{lastpage}} \pubyear{}

\maketitle

\label{firstpage}

\begin{abstract}
As part of a large spectroscopic survey of $z>5$ Lyman break galaxies (LBGs), we have identified a single source which is  clearly hosting an AGN. Out of a sample of more than fifty spectroscopically-confirmed $R-$band dropout galaxies at $z\sim5$ and above, only J104048.6-115550.2 at $z=5.44$ shows evidence for a high ionisation-potential emission-line indicating the presence of a hard ionising continuum from an AGN. Like most objects in our sample the rest-frame-UV spectrum shows the UV continuum breaking across a Ly$\alpha$ line. Uniquely within this sample of LBGs, emission from NV is also  detected, a clear signature of AGN photo-ionisation. The object is spatially resolved in HST imaging. This, and the comparatively high Ly$\alpha$/NV flux ratio indicates that the majority  of the Ly$\alpha$ (and the UV continuum longward of it) originates from stellar photo-ionisation, a product of the ongoing starburst in the Lyman break galaxy. Even without the AGN emission, this object would have been photometrically-selected and spectroscopically-confirmed as a Lyman break in our survey. The measured optical flux ($I_{AB}=26.1$) is therefore an upper limit to that from the AGN and is of order 100 times fainter than the majority of known quasars at these redshifts. The detection of a single object in our survey volume is consistent with the best current models of high redshift AGN luminosity function, providing a substantial fraction of such AGN are found within luminous starbursting galaxies. We discuss the cosmological implications of this discovery.

\end{abstract}


\section{Introduction}
\label{sec:intro}

Recent years have seen rapid observational progress at $z>5$. `Dropout' surveys, searching for the optical signature of absorption by the dense high redshift intergalactic medium (IGM), have revealed a population of star-forming galaxies within 1\,Gyr of the Big Bang \citep[e.g.][]{2003ApJ...593..630L,2003MNRAS.342..439S,2004ApJ...600L..99D,2004ApJ...606L..25B}. At the same time, the huge but shallow Sloan Digital Sky Survey \citep[SDSS, ][]{2000AJ....120.1579Y} has revealed a population of highly luminous quasars, extending as far back as $z=6.4$ \citep{2006AJ....131.1203F}.

Observations of galaxies and Active Galactic Nuclei (AGN) at these high redshifts give us a unique window into the early stages of structure formation in the universe, and on the effects that the first stars and galaxies had on their surroundings. Analysis of fossil populations in nearby galaxies and simulations of hierarchical galaxy formation both predict that the galaxies observed at $z>5$ are the progenitors that will eventually merge to form today's massive galaxies and clusters $\it{e.g.}$ \citet{th2005}. The mass accumulation and merger process can be studied over cosmic time, feeding back into cosmological simulations.

Despite the success of recent observational work, large gaps still remain in our understanding of galaxies and quasars at this epoch and in particular how they relate to each another. Neither the population of bright galaxies nor faint AGN have been well constrained at $z>5$. Not only are the populations expected to overlap at faint magnitudes and at low space densities, both populations can be targeted by the same large-area deep optical surveys. A current topic of much debate is the influence of AGN activity on star formation within a galaxy and vice versa. Many galaxy formation models invoke AGN feedback to quench star formation and prevent the formation of many small galaxies which are over predicted by dark matter simulations. The details of the form of this AGN feedback are still unknown. What work that has been done on the AGN fraction in samples of $z>5$ Lyman break galaxies (LBGs) has indicated that this fraction is small. For example \citet{bremer04} showed that none of a sample of over 50 photometrically-selected $z>5$ LBGs showed any evidence of X-ray emission down to deep limits ruling out strong hidden AGN components in them. \citet{2003MNRAS.342..439S} obtained similar null results for $z\sim6$ LBGs. A second X-ray luminosity contraint is from the stacking anyalsis of \citet{lehmer05} who used a larger sample of galaxies within the same field as used by \citet{bremer04,2003MNRAS.342..439S}. \citet{lehmer05} place X-ray luminosity limits of $2.8x10^{41}$ and $7.1x10^{41} erg s^{-1}$ for $z>5$ and $z>6$ LBGs respectively.

This is consistent with the results of observations at lower redshifts. \citet{2006AJ....131...49R} and \citet{2001AJ....122.2833F} have inferred a low number density and contribution to reionisation by AGN from the bright end slope of the $z>4$ AGN luminosity function. This result has been supported by \citet{2005ApJ...634L...9M} who reported the detection of just one faint high redshift quasar in a sample surveying 2.5\,deg$^2$ and selecting for flat-spectrum optical sources. Nonetheless, none of these observations have probed below the knee in the tentative $z>5$ quasar luminosity function and hence the current observations of quasars with M$_{1450}<-27$ do not preclude the existence of faint AGN which could reasonably contribute a significant fraction of the ionising and ultraviolet luminosity density at high redshift.

In this paper we present the photometric and spectroscopic properties of an AGN detected at extremely faint magnitudes in a large multiwavelength survey. In section \ref{sec:observations} we briefly describe the survey and present the source properties. In section \ref{sec:discussion} we consider the implications of this detection for our understanding of high redshift surveys and the quasar luminosity function. Finally in section \ref{sec:conclusions} we summarise our conclusions.

We adopt the standard Lambda cosmology, i.e. a flat universe with $\Omega_{\Lambda}=0.7$, $\Omega_{M}=0.3$ and $H_{0}=70 h_{70} {\rm km\,s}^{-1}\,{\rm Mpc}^{-1}$. All magnitudes are quoted in the physically meaningful AB system (Oke \& Gunn 1983) in which a source flat in $f_\nu$ has zero colour.

\section{Observations and Object Selection}
\label{sec:observations}

The source J104048.6-115550.2, was photometrically selected and spectroscopically followed-up as part of a large sample of $z>4.8$ Lyman break galaxies. The sample was selected from multiband imaging with VLT/FORS2 (two hours in each of $V$, $R$, $I$ and $z$ bands) and NTT/SOFI (five hours in $J$ and six hours in $K_{s}$) of ten widely-separated fields covering a total sky area of about 430 sq arcminutes. All images except the $z$-band were taken as part of the EDisCS survey \citep{2005A&A...444..365W}. The details of sample selection will be presented elsewhere (Douglas et~al. 2007, in preparation), we briefly note here how the source was selected.  

The source was photometrically selected to be a $z\sim5$ LBG candidate, with a non-detection in $V$, a colour cut of $R-I>1.3$ and non-detection in $J$ and $K_{s}$ ($J> 24.9, K_s>23.9$). The latter near-IR requirement rejects interloping $z\sim 1$ galaxies (which can have similar optical colours) and many cool stellar (or substellar) objects that can contaminate LBG samples. In fact this colour selection will not only identify LBG candidates but also any unobscured AGN at similar redshifts given that they have similar colours. As with all other candidates, photometry was initially carried out using Sextractor \citep{1996A&AS..117..393B} on each of our optical images.  The photometry for this object was then remeasured with IDL routines using $2''$ apertures centered on the $I$-band position and the results were indistinguishable from those measured by Sextractor. Photometric limits and uncertainties were measured from the standard deviation of the background in apertures of the same size. The flux was measured in randomly placed apertures and background levels were subtracted using the corresponding smoothed background used by Sextractor in its calculations. This method exactly matches the photometric measurements, thus providing depths which are closely related to the photometry. The photometric properties of J104048.6-115550.2 and $2\sigma$ depths for each band are quoted in Table \ref{photom}.

\begin{table}
\caption{Photometric properties of J104048.6-115550.2 and 2$\sigma$ photometric depths of images measured in $2''$ apertures}
\label{photom}
\begin{tabular}{ccccccc}
\hline
 & $V$ & $R$ & $I$ & $z$ & $J$ & $Ks$ \\ \hline
Aper& $>27.8$ & $>27.5$ & $25.9\pm0.3$ & $25.8\pm0.3$ & $>24.9$ & $>23.9$ \\
Corr& $>27.9$ & $>27.6$ & $26.1\pm0.3$ & $26.0\pm0.3$ & $>25.1$ & $>24.1$ \\
M$_\mathrm{lim}$& 27.8 & 27.5 & 26.7 &26.3 &24.9 & 23.9 \\
\hline
\end{tabular}

\medskip
Row aper is the photometry of the object within $2''$ diameter apertures with 1$\sigma$ errors. Row corr is the photometry after galactic extinction ($<0.15$ magnitudes) and lensing from foreground objects (0.25 magnitudes) corrections using the model of \citet{clowe} were applied. The third row corresponds to the 2$\sigma$ photometric depths of images measured in $2''$ apertures without extinction of lensing corrections.
\end{table}

This object and other LBG candidates were spectroscopically observed on FORS2 VLT/UT1 on the $1^\mathrm{st}$ and $4^\mathrm{th}$ May 2005. The candidates were observed in `MXU' mode with the 300I grating and the OG590 blocking filter centered on a wavelength of 7600{\AA}. With a slit width of $1''$, this set up produced a dispersion of 3.2{\AA} per pixel across the wavelength scale and a spatial scale of $0.2''$ per pixel. Twenty spectroscopic frames were taken of the target, each for 650 seconds. The object was placed at a different pixel position along the slit for each exposure to facilitate sky subtraction and to avoid bad pixels. The data were then reduced following the prescription in \citet{2003ApJ...593..630L}. Each frame was bias subtracted and flat fielded. The sky was then subtracted by producing a sky frame from the three observations closest in time. To improve the skyline residuals a first-order polynomial fit to each column was also subtracted. The spectra were then shifted and combined, wavelength calibrated using a lamp observations and flux calibrated using the spectrophotometric standard Feige 56.
 
Of the dozens of spectra of LBGs obtained as part of the survey, some tens of spectra show a clear Ly$\alpha$ line coupled with a continuum break and no other clearly identified spectral features (these spectra will be presented in Douglas et~al. 2007). Only J104048.6-115550.2 additionally shows a clear, but faint NV line. Lines as faint as this have been detected before in distant LBGs, but have been identified as Ly$\alpha$ \citep[][Stanway et~al. 2007]{2003ApJ...593..630L}. The line does not fall on any obvious strong sky features. 

The source was spectroscopically observed in a comparable way to the infrared spectroscopic technique of observing in a ABBA repeated offset pattern resulting in a final spectrum with a positive and negative signal. Instead of one negative signal our adapted technique results in numerous negative signals in the sky background at different spatial positions. This is because of the way the sky frames are assembled from several data frames nodded along the slit and then subtracted from another of the data frames. Normally the Ly$\alpha$ and continuum emission from our objects are sufficiently faint that the negative artifacts in the sky background do not affect the final object spectrum. However, the Ly$\alpha$ line in this object is comparatively strong and consequently the negative features produced by the sky-subtraction procedure increases the uncertainty in determining the Ly$\alpha$ line flux and line shape. To minimise this effect the data were rereduced following a different perscription. 

\begin{figure*}
\begin{center}
\includegraphics[width=17cm]{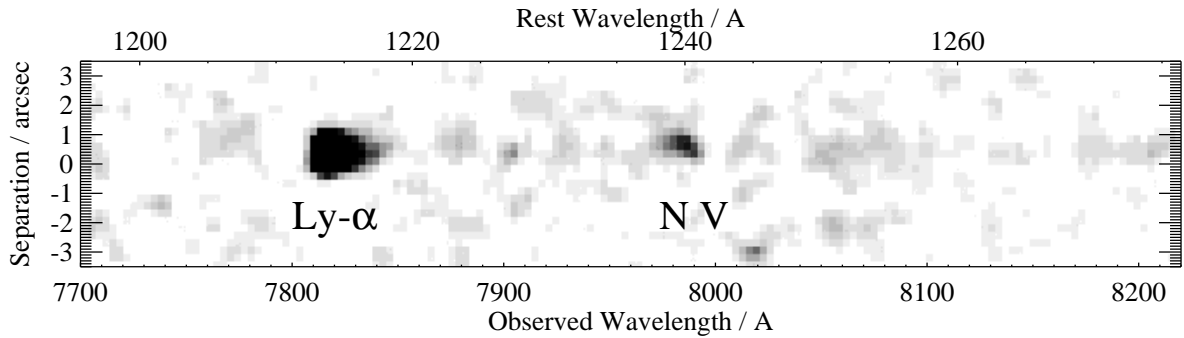}
\caption{FORS2/VLT two dimensional spectrum of J104048.6-115550.2 clearly showing the typically asymmetric profile of a Ly$\alpha$ line with a sharp blue cutoff and a extended red wing, and the detection of NV.}
\label{2dspec}
\end{center}
\end{figure*}

The bias subtracted and flat fielded images were each sky subtracted by fitting a second order polynomial to each column (unique wavelength bin) of the data using an IDL procedure. Cosmic rays were excluded using sigma clipping (2$\sigma$). Each sky subtracted frame was shifted so the object spectrum fell on the same spatial pixel. The final image was made from the average of these frames rejecting the two highest and lowest values out of 20 samples at each position thus rejecting cosmic rays and bad pixels. Figures \ref{2dspec} present this reduction of the data which is comparable to the initial reduction, apart from the improvement in the characterisation of Ly$\alpha$. The Ly$\alpha$ line has the truncated, asymmetric shape expected of a high redshift Ly$\alpha$ line suffering absorption on the blue side from the Ly$\alpha$ forest. The NV is clearly visible but at a lower flux than the Ly$\alpha$ (fainter by a factor of $\sim$ 13).

\begin{table}
\caption{Spectroscopic line properties of J104048.6-115550.2}
\label{specprop}
\begin{tabular}{ccccc}
\hline
Line & $\lambda_{obs}$ / \AA & $\lambda_{rest}$ / \AA & Flux / ergs cm$^2$ s$^{-1}$ & $W_{0}$ / \AA \\
\hline
Ly$\alpha$ & 7829 & 1216 & $3.23\pm0.65 \times 10^{-17}$& $127\pm45$\\
NV & 7989 & 1240 & $2.43\pm0.73 \times 10^{-18}$  & $10\pm4$ \\
\hline
\end{tabular}
\end{table}

\section{Analysis and Discussion}
\label{sec:discussion}

\subsection{AGN vs Starburst Contribution}

The broad band colours of this source are consistent with those of a starbursting host galaxy (flat in $f_{\nu}$ and modified by IGM absorption), undetected in $V$ and $R$ bands and flat in colour between the $I$ and $z$ bands. However, given the indications of AGN activity, what fraction of the light observed is from the faint AGN? The fact that NV has been observed means that at least several tens of percent of the light must be attributed to the AGN. This also agrees with the Ly$\alpha$/NV ratio observed. Through theoretical calculations and observations it has been found that in a typical quasar the Ly$\alpha$/NV ratio is between four and seven \citep{oster,vanden01,dietrich,kurasz04}. This object has a Ly$\alpha$/NV ratio of 13.3, a factor of 2-3 higher than that expected. This can be explained if there is large contribution to the Ly$\alpha$ flux from the star formation in the host galaxy. Assuming the typical quasar ratio, the AGN is contributing 30-50\% of the total flux of the source making it $0.5-2$ magnitudes fainter than that quoted. Of course this is only an estimate due to uncertainties in the Ly$\alpha$/NV ratio measurements. Due to the uncertainties in the broad component measurement of the Ly$\alpha$ line caused by the low resolution of the spectrum we are unable to determine the effects of opening angle which could potentially brighten the AGN.
 
In addition to the multiband ground-based imaging this object has also been observed by the Hubble Space Telescope. The field containing J104048.6-115550.2 was observed by the ACS in the F775W filter for four orbits in the center of the field and for one orbit around the edge.  The half-light radius measured for J104048.6-115550.2 in the one orbit area was 0.095$''$. To test whether this is consistent with the object being spatially resolved, the PSF, as measured from a 23 magnitude star in the four orbit region, was faintened to I$_{AB}$=26 and placed randomly on the HST image in areas only covered by one orbit. It was found that only 5\% of recovered objects had a half-light radius greater than 0.09$''$ with a mean radius of 0.07$''$, implying that this source is resolved. If a point source profile of the same brightness is added to the object image, the resulting half-light radius drops to within the unresolved range showing that it is unlikely that this object has an AGN contributing more than 50\% of the total flux.

An alternative explanation of the high Ly$\alpha$/NV ratio is a lower metallicity within the nucleus of the galaxy. The expected Ly$\alpha$/NV ratio is taken from bright QSOs with solar or supersolar metallicities in the nucleus. A metallicity lower than solar concentrations would explain the weak NV emission. However from studying figure six in \citet{hamann99}, the metallicity of the core would have to be a hundredth of solar to account for the line ratio. \citet{hamann99} also show that from elemental abundances in AGN at low and high redshift out to $z>4$, the gas in AGN appears to have a solar to supersolar metalicity. \citet{verma} found consistent fits to the SEDs of $z\sim5$ LBGs assuming 0.2 Z$_{\odot}$. In addition, \citet{verma} suggest that the typical $z\sim5$ galaxy is young, only a few 10 Myrs old, and drive vigorous starburst-driven outflows.  Such constraints would also suggest relatively low chemical abundance in the typical LBGs at $z\sim5$. These results would make the galaxy significantly more metal rich than the nucleus, a situation highly unlikely given the small mass in the nuclear region compared to the galaxy. Assuming that the high Ly$\alpha$/NV ratio is due to the contribution of star-formation suggests that the `narrow line regions' ($\it{i.e.}$, within about 10pc of the AGN) has already reached solar or super solar metallicities at z=5.4.

The absolute magnitude of this object at 1400{\AA} is -22.5. However as the host galaxy contributes a large fraction of the observed light the magnitude of the AGN is likely to be between -21 and -22 making this the faintest AGN seen at $z>5$. Although the Sloan Digital Sky Survey and others \citep{stern2000,rom2004,2005ApJ...634L...9M} have found QSO at similar and higher redshifts they are all at least two magnitudes brighter. At this faint level in the rest frame UV, the host galaxy luminosity function is overlapping with the faint end of the QSO luminosity function.

In detailed photometric and spectroscopic work at $z\sim3$ \citet{2003ApJ...592..728S} found a AGN fraction of 3\% in their comparable LBG sample. This was confirmed by the study of the X-ray properties of the same $z\sim3$ sample \citep{laird06}. From the one AGN detected in this $z>5$ sample the AGN fraction of $\sim2\%$ agrees with that found at $z\sim3$. However, studies comparing the rest-frame optical and UV properties of LBGs observed at $z\sim3$ and $z\sim5$ have found significant differences between the two populations. By using the same selection and analysis used in the $z\sim3$ sample, \citet{verma} have found that $z\sim5$ galaxies are typically much younger and have lower stellar masses than their $z\sim3$ counterparts, demonstrating a clear change in properties of the LBG between $z\sim3$ and $z\sim5$. With evidence of evolution in stellar age and mass, changes in other properties such as AGN fraction would be expected though surprisingly not seen in this large spectroscopically-confirmed sample. 

A comparable spectroscopic study by \citet{vanzella} in the GOODS CDFS field have not reported on the AGN fraction to date although they have not completed their study so cannot yet add to the statistics presented here. They also spectroscopically followed-up photometrically-selected $z>5$ LBG candidates. We encourage other authors to closely investigate spectroscopic follow-up of such objects and report positive and negative detections of AGN activity to provide more comprehensive statistics in this field.

\subsection{Velocity Profile of Line Emission}

\begin{figure}
\begin{center}
\includegraphics[width=8cm]{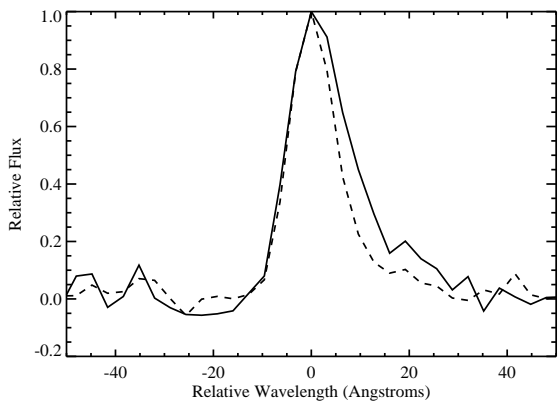}
\caption{Ly$\alpha$ line profile for average line emitters (dashed line) and J104048.6-115550.2 (solid line). The X-axis is measures in observed wavelength relative to the peak of the line. Neither of the curves is deconvolved from the instrumental response because the average of the eight appear to be unresolved.} \label{hst}
\end{center}
\end{figure}

The presence of a clear NV line strongly suggests that this galaxy, selected through its star formation properties, also contains a significant AGN component. As Figure \ref{hst} illustrates, the Ly$\alpha$ line in the AGN candidate is one of the strongest in the spectroscopic sample and is the broadest of all lines of comparable flux. The dashed line in this figure shows the average profile of the 8 Ly$\alpha$ emitters with similar line fluxes (and luminosity) to the main object.  All 8 Ly$\alpha$ line are unresolved by the spectrograph setup, some with indications of an asymmetric red wing. The average measured FWHM of the sample (omitting the AGN candidate) is about 400 km/s, the resolution of the spectroscopic setup. 

The emission line in J104048.6-115550.2 matches those from star forming galaxies in the blue slope (consistent with a sharp drop due to absorption), but has a clear red wing. It is resolved, with a measured FWHM of about 500 km/s which, when deconvolved with the instrumental response, relates to $\sim300$ km/s. However, if we assume much of the Ly$\alpha$ emission comes from an unresolved stellar-photoionised component, the AGN component is broader. Subtracting 50 per cent of the line using an unresolved component broadens the remaining flux to around 1000 km/s, although this figure is indicative rather than precise as the star formation contribution is uncertain. 

Although objects with strong, wider Ly$\alpha$ emission are more likely to have AGN components contributing to the flux, the other eight sources with similar Ly$\alpha$ flux and comparable equivalent widths show no evidence of an AGN component, demonstrating that such large Ly$\alpha$ lines can be explained by star formation alone. Apart from the NV line, J104048.6-115550.2 does not exhibit other properties which are extreme for a $z>5$ LBG.

This broad Ly$\alpha$ component and the presence of NV, only seen through AGN activity, confirms the identification of a central AGN. As such this is the faintest $z>5$ AGN seen to date by at least two magnitudes. Out of the tens of spectroscopically confirmed Lyman break and Ly$\alpha$ emitting galaxies in our ten survey fields only this object, J104048.6-115550.2, shows any NV emission suggesting that although these objects are present at this redshift epoch they are very rare. It is likely that some of the other Ly$\alpha$ emitters may have very weak AGN which are being swamped by the vigorous ongoing star formation in the host galaxies. We would not have detected NV in such objects.

\subsection{Implications for the AGN Luminosity Function}

The QSO luminosity function is fundamental to cosmology. It is intimately linked to many aspects of cosmology that have yet to be fully characterised, such as galaxy and supermassive black hole formation, the accretion history of black holes and the role of galaxy mergers and interactions in galaxy formation and evolution. The Sloan Digital Sky Survey has provided good constraints on the bright end of the luminosity function but can not probe the low luminosity end occupied by Seyfert type AGN at high redshift. To probe these faint levels a photometric and spectroscopic survey at levels more normally used to search for LBGs, such as the work described here, is needed. The faint-end slope has been the topic of some debate as there is growing evidence that it is flattening with increasing redshift, seen in the X-ray by \citet{ueda} and \citet{franca}. The high-z faint-end slope is important in determining the early formation history of black holes and their impact on galaxy evolution. 

From a survey covering 430\,arcmin$^{2}$ with over 85\% spectroscopic completeness of the photometrically selected candidates plus additional spectra of less robust candidates, J104048.6-115550.2 was the only object to show evidence for AGN activity. As the photometric selection criteria of AGN and LBGS is the same at these redshifts the uncertainity of our one AGN detection out of 50 confirmed LBG candidates is around a factor of two. At this flux level the NV line could potentially have been detected in any of the spectra. Therefore at M$_{1400}\approx -21$ the surface density of AGN observed in this survey is 2.3 x 10$^{-3}$ arcmin$^{2}$.

\begin{figure}
\begin{center}
\includegraphics[width=8cm]{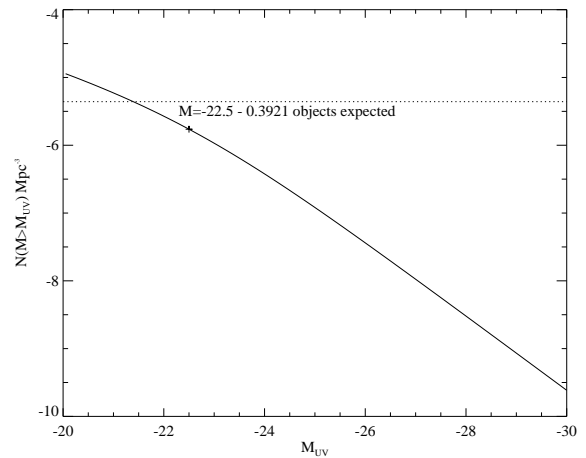}
\caption{Luminosity function of QSO using \citet{2006AJ....131.2766R} and \citet{2006ApJ...639..700H} for the faint-end slope. The horizontal dotted line shows the number density corresponding to one AGN detected in our survey area.} \label{LF}
\end{center}
\end{figure}

The QSO luminosity function of \citet{2006AJ....131.2766R} modified by the theoretical faint-end slope of \citet{2006ApJ...639..700H} is shown in Figure \ref{LF}, calculated for $z=5$. \citet{2006ApJ...639..700H} show that the evolution of the faint-end slope seen at lower redshifts can be explained by luminosity-dependent quasar lifetimes and the physics of quasar feedback, where the faint-end slope is explained by quasars with a peak luminosity near the luminosity function break but seen in a less luminous stage of evolution. This is an alternative to assuming there are changes in the distribution of the peak luminosity of quasars and of the masses of the black holes. The detailed modelling of the faint-end slope by \citet{2006ApJ...639..700H} is motivated by its ability to give insight into the distributions of central black hole masses, host properties and quasar fuelling mechanisms and accretion. Therefore the faint-end of the AGN luminosity function is a sensitive probe of the formation of structure at high redshift. 

At the luminosities probed by this survey, the faint-end slope is beginning to significantly modify the expected number of objects predicted. In our survey we would expect to find 0.4 objects with AGN activity at M=-22.5, the total magnitude of this object. Given that the AGN emission accounts for less that 50\% of the UV luminosity of this object the AGN is intrinsically fainter than this by up to a magnitude. The modified luminosity function predicts one object in our survey volume at this fainter luminosity. Consequently our identification of a single AGN agrees with the theoretical predictions provided that all AGN at this redshift are in starbursting LBGs. 

If there are additional AGN at this lower flux level not embedded in a UV luminous LBG so appearing as a `naked' nucleus, these would not have been selected within our survey as they would be fainter than our $I$-band flux limit. If this were the case we would expect to see similar `naked' nuclei at brighter magnitudes which would have been spectroscopically identified as AGN in our survey. Given the slope of the AGN luminosity function, our nondetection of bright `naked' AGN requires that a significant fraction of all faint AGN are within luminous LBGs.

The prospect of identifying a large sample of faint AGN in order to constrain the faint-end of the luminosity function is daunting. Not only would such surveys require an area of many square degrees at a comparable depth to ours in optical and IR bands, but disentangling AGN and starburst emission in such sources is limited by the sensitivity of the deepest spectroscopic observations currently possible. To do so cleanly requires characterisation of multiple metal lines in spectra. This is likely to require the capabilities of future large telescopes either space based (JWST) or on the ground (ELTs).

\subsection{Cosmological Implications}

The detection of a single AGN in a sample of 50 or more UV starbursts indicates that faint AGN contribute an insignificant amount of the meta-galactic ionising flux at $z>5$. It is already known that brighter SDSS quasars \citep{2006AJ....131.1203F} contribute less ionising flux than the emsamble of starbursting galaxies at $z>5$. Until now there was the possibility that a large population of faint AGN could rival the starbursts for UV output. This is clearly ruled out. Indeed the fraction of ionising flux provided by AGN decreases as a function of luminosity. With one source out of 50 containing an AGN which provides at most 50\% of its flux, only 1\% of ionising photons emitted by $M_{UV}\sim-22$ is provided by AGN. Given that we are observing these objects just after reionisation the obvious conclusion is that AGN have no global influence on the ionisation state of the IGM.

The majority of LBGs at $z\sim5$ appear to be undergoing their first generation or generations of star formation typically with stellar populations younger than 100 million years \citep{verma}. The stellar density of these systems is only matched at low redshift in the centers of the massive galaxies indicating that these sources will eventually evolve into the central regions of such galaxies today. This is supported by analysis of the spectra of near-by galaxies which indicates that the most massive and dense systems formed their stars at these redshifts ($\it{e.g.}$ \citet{th2005, heavens}). It is perhaps not surprising that we have identified a massive black hole in at least one galaxy given that these objects are routinely found at the center of massive galaxies. In at least one instance we are seeing simultaneous AGN and galaxy formation activity, indicating at least at this luminosity AGN activity does not quench star formation. 

\section{Conclusions}
\label{sec:conclusions}

In a spectroscopic sample of LBG at $z\sim5$ we have found a single object showing clear signatures of AGN emission. Analysis of its spectrum and HST imaging indicates that no more than 50\% of the UV flux arises from the AGN, the rest comes from the starburst in the host LBG. 

This single detection is consistent with the predictions of the best available luminosity function for faint AGN only if a significant fraction of all faint AGN are found in strongly starbursting galaxies. 

Disentangling the AGN emission from the starburst emission will be a major technical challenge in future observational studies of the faint-end of the luminosity function at these redshifts.

At these flux levels the contribution of AGN to the total ionising photon budget at $z\sim5$ is of order of 1\%. Given the short time elasped since the end of reionisation, this implies that the role of AGN in the global reionisation of the universe is small.

\section*{Acknowledgments}

We thank the referee for their helpful comments. LSD and ERS acknowledge support from PPARC. Based on observations made with ESO Telescopes at the La Silla and Paranal Observatory under programme IDs 166.A-0162 and 175.A-0706. Based on observations made with the NASA/ESA Hubble Space Telescope, obtained [from the Data Archive] at the Space Telescope Science Institute, which is operated by the Association of Universities for Research in Astronomy, Inc., under NASA contract NAS 5-26555. These observations are associated with program 9476.

\label{lastpage}

\end{document}